\documentclass[cameraready]{Interspeech}

\title{UD-ASD: A Unified Diffusion Model for Anomalous Sound Detection}

\author[affiliation={1}, equalcontribution]{Pengxiang}{Gao}
\author[affiliation={1}, equalcontribution]{Yu}{Qiu}
\author[affiliation={1}, orcid=0000-0000-0000-1111, correspondingauthor]{Yanzhi}{Song}

\address{
    $^1$ University of Science and Technology of China, China
}

\email{pxgao@mail.ustc.edu.cn, 1481082109@qq.com, yanzhis@ustc.edu.cn}


\usepackage{comment}
\usepackage{multirow}

\begin{document}

\maketitle

\begin{abstract}

    
    Anomalous Sound Detection (ASD) aims to determine whether faults have occurred by monitoring sounds. Existing methods detect a limited range of anomalies, exhibit poor generalization, or train a separate model for each machine. Diffusion models possess strong generalization and can generate specific data with condition guidance. We propose a unified diffusion model only with a small module. The audio is first transformed into log-Mel spectrograms. The lightweight module embeds machine IDs into condition embeddings, guiding the model to reconstruct data for specific machines. Then diffusion model reconstructs data with condition, using Gaussian Mixture Models to fit the distributions of reconstruction errors. Our unified model could monitor multiple machine types and learn more fundamental feature spaces with cross-domain learning. Experiments on DCASE2022 Challenge Task 2 show that our model achieves 3.44\% AUC and 2.52\% pAUC improvements over baseline, validating its effectiveness.
\end{abstract}

\section{Introduction}

Anomalous Sound Detection (ASD) involves analyzing the operating sounds of machinery to identify potential malfunctions, providing non-contact, cost-effective, and efficient real-time monitoring \cite{FMQAP_2025}. The fundamental principle of ASD is to model the distribution of normal audio, such that any deviation from this learned distribution is flagged as an anomaly \cite{DOSCL_2025, AASRL_2025}.

Existing ASD methods can be broadly categorized into three main types. First, traditional methods rely on hand-crafted features extracted from time-domain \cite{kanai_2023} (e.g., amplitude, energy) or frequency-domain \cite{guan_2023} (e.g., Fourier transform) representations of the audio signal \cite{vafeiadis_2020}. While some approaches incorporate neural network-based audio features \cite{dpNICE_2025}, these methods generally depend heavily on domain-specific feature engineering, which increases annotation costs and limits their generalization capabilities. Second, self-supervised learning methods \cite{GLAM_2025, clp-scf_2023, stgram_2022} leverage auxiliary tasks, often guided by meta-labels (e.g., machine ID, operational conditions), to learn a discriminative representation of normal audio. The goal is to create a feature space where anomalous sounds are clear outliers \cite{DOSCL_2025, AASRL_2025, hadi_2022_icassp, AnoPatch_2024}. However, their performance can be sensitive to ambient noise, and the reliance on auxiliary labels restricts their applicability in scenarios where such metadata is unavailable.

The third category, generative methods, learns to reconstruct normal audio with high fidelity. When presented with an anomalous audio, these models produce a high reconstruction error, which serves as the anomaly score. Common approaches include Autoencoders (AEs) \cite{zhang_2025_dsp, AEGAN-AD} and, more recently, denoising diffusion models \cite{ASD-Diffusion_2024, ssdm_2024, ASD-Diff-fan24}. While these methods reduce the dependency on annotations, they have significant drawbacks. AE-based models often suffer from poor generalization and may degenerate into learning identity functions \cite{Remove-Anomalies}. More critically, existing generative approaches, both AE- and diffusion-based, typically require training a separate model for each machine type. This one-model-per-machine paradigm leads to prohibitive storage and training costs in real-world, multi-machine environments. When multiple machines speak together, these methods require training multiple models to monitor for anomalies. 

To address these limitations, we propose UD-ASD, a \textbf{U}nified \textbf{D}iffusion model for \textbf{A}nomalous \textbf{S}ound \textbf{D}etection, which employs a unified conditional diffusion model to reconstruct spectrograms and detect anomalies with Gaussian Mixture Model (GMM) for all machines. Our framework consists of three main components: a Condition Projector (CP), a conditional diffusion model, and a Gaussian Mixture Model (GMM) for anomaly scoring. 
The CP first encodes machine ID information into a condition embedding. This embedding is concatenated with the input log-Mel spectrogram, providing the necessary context to the diffusion model. The model is trained exclusively on normal data to perform a denoising-based reconstruction task. During inference, the model attempts to reconstruct any given input as a normal sound. Consequently, anomalous inputs result in a high reconstruction error. To account for temporal context, we apply temporal pooling to the frame-wise reconstruction errors. Finally, a GMM, also trained only on the reconstruction errors of normal data, models this error distribution. The final anomaly score is derived from the negative log-likelihood of a given sample's error, effectively measuring its deviation from the norm. 
Our contributions are: 1) We propose a unified diffusion framework that reduces training overhead by eliminating the one-model-per-machine constraint. 2) We introduce Condition Projector, a lightweight module provides condition information guidance for the framework. 3) We achieve sota performance on the DCASE2022 Task 2 dataset \cite{dataset}, demonstrating both superior accuracy and efficiency. The code will be open-source after the review process.

\section{Proposed method}\label{sec:method}

\begin{figure}[t]
    \centering
    \includegraphics[width=0.9\linewidth]{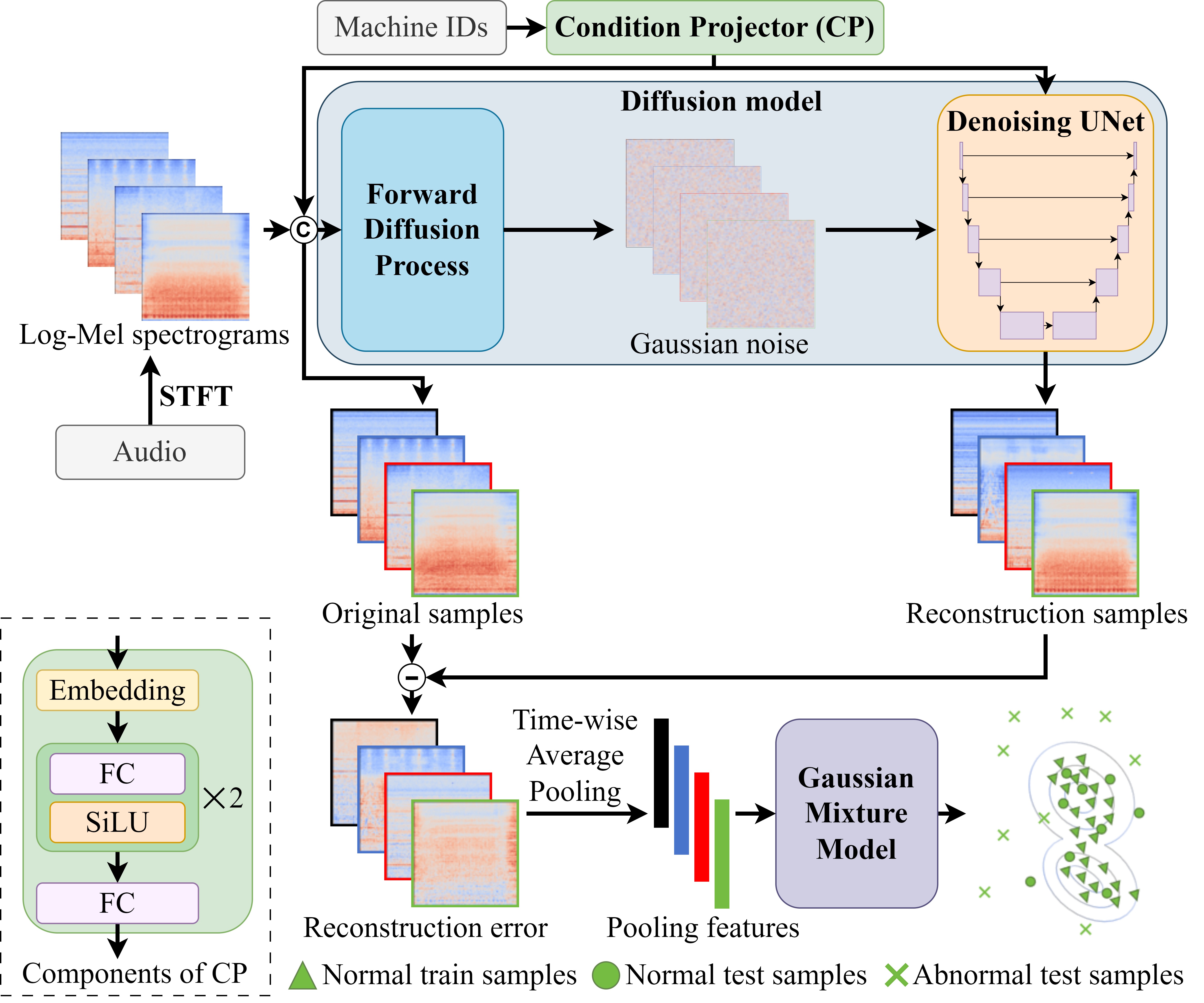}
    \caption{Overview of UD-ASD. Different colors indicate various machines. Diffusion model includes forward diffusion and denoising UNet. UD-ASD is trained only on normal data for lower reconstructed error. During testing, the mean of error is calculated across time-dim and is input into GMM to detect.}
    \label{fig:UD-ASD}
\end{figure}

The architecture of our proposed UD-ASD model is illustrated in Figure~\ref{fig:UD-ASD}. The framework operates in three main stages. First, input audio is transformed into a log-Mel spectrogram via the Short-Time Fourier Transform (STFT). Concurrently, a CP module encodes the machine's ID into a condition embedding, which is concatenated with the corresponding spectrogram. Second, a conditional Diffusion Model, trained exclusively on normal data, takes the conditional data as input to perform a reconstruction task. During inference, this process yields a high reconstruction error for anomalous inputs. Finally, an Anomaly Scoring system, based on a GMM, models the distribution of reconstruction errors from normal data and calculates the final anomaly score for a given test sample.

\subsection{Condition projector}\label{sec:conditionemb}

Our primary innovation directly addresses that one-model-per-machine limitation inherent in most existing generative ASD methods, simple yet significantly reducing deployment overhead. We introduce the Condition Projector, a lightweight module shown in Figure~\ref{fig:UD-ASD}, to map discrete machine identity labels $c$ (e.g., machine type or ID) into dense condition embeddings.

In UD-ASD, different machine IDs $c$ are encoded as condition embeddings with CP. The embeddings are combined with the spectrograms and passed through each layer of UD-ASD to ensure full integration of condition information. Diffusion models adds noise to the original clear spectrograms $\mathbf{x}^{org}_0$ with $t$-th step, resulting in the corrupted data $\mathbf{x}^{org}_t$.
\begin{align}\label{eq: conditional_diffusion}
    \mathbf{x}_t = {\rm concat} (\mathbf{x}^{org}_t, {\rm CP}(c)) \ ,
\end{align}
where the ${\mathbf x}_t$ is the $t$-th input with condition and $\mathbf{x}^{org}_t$ is the $t$-th original data without condition ($t \in [0, 1, ..., T]$). For convenience, subsequent instances of ${\mathbf x}_t$ denote conditional data. The ${\rm concat} (\cdot)$ means channel concatenation. ${\rm CP}(\cdot)$ denotes CP module, whose structure is shown in the lower left of Figure~\ref{fig:UD-ASD}. CP maps discrete labels to embeddings, allowing UD-ASD to generate machine-specific data with additional information.

\subsection{Diffusion model}\label{sec:diffusion}

Our reconstruction backbone is a conditional diffusion model, trained using the Denoising Diffusion Probabilistic Model (DDPM) framework \cite{ddpm} (the forward diffusion process $q(\mathbf{x}_t \, | \, \mathbf{x}_{t-1})$ and the reverse denoising process $p_{\theta}(\mathbf{x}_{t-1} \, | \, \mathbf{x}_t)$) and accelerated during inference using the Denoising Diffusion Implicit Model (DDIM) sampler \cite{ddim}. 

\textbf{In the forward diffusion process}, the corrupted data $\mathbf{x}_t$ in the $t$-th step is generated with a Markov chain by adding Gaussian noise to the original input data $\mathbf{x}_0$ step by step: 
\begin{align}
\label{eq: forward_process_neibor}
    q(\mathbf{x}_t \, | \, \mathbf{x}_{t-1}) = \mathcal{N}(\mathbf{x}_t \, | \, \sqrt{{\alpha}_t}\mathbf{x}_{t-1}, \ (1 - {\alpha}_t)\mathbf{I}) \ ,
\end{align}
and $\alpha_t$ represents the noise variance schedule used to regulate the noise added at the $t$-th step. As the noise schedule we employ in the forward diffusion is a linear Gaussian process, the closed-form solution can be directly computed from $\mathbf{x}_0$: 
\begin{align}
\label{eq: forward_process}
    q(\mathbf{x}_t \, | \, \mathbf{x}_{0}) = \mathcal{N}(\mathbf{x}_t \, | \, \sqrt{\bar{\alpha}_t}\mathbf{x}_{0}, \ (1 - \bar{\alpha}_t)\mathbf{I}) \ ,
\end{align}
where $\bar{\alpha}_t = \displaystyle \prod_{i=1}^{t} \alpha_i$. So the corrupted data $\mathbf{x}_t$ could be computed based on the original input data $\mathbf{x}_0$ and the $t$-th step: 
\begin{align}
\label{eq: forward_proces_x}
\begin{aligned}
    \mathbf{x}_t = \sqrt{\bar{\alpha}_t} \ \mathbf{x}_0 + \sqrt{1 - \bar{\alpha}_t} \ \epsilon_t \ ,
\end{aligned}
\end{align}

In the actual model, Eq.~(\ref{eq: forward_proces_x}) is used to generate the pure Gaussian noise from the original conditional log-Mel spectrograms $\mathbf{x}_0$ with the noise schedule and the random noise $\epsilon_t$ generated at step $t$.

\textbf{Reverse process and training.} In the reverse denoising process, the corrupted data $\mathbf{x}_t$ are progressively denoised to noise-free data $\mathbf{\tilde{x}}_0$ by predicting the real Gaussian noise $\epsilon_t$. We employ a UNet-like attention-based network \cite{beatgan} $\mu_{\theta}(\cdot, \cdot)$ with parameters $\theta$ to predict $\epsilon_t$. $\mu_{\theta}(\cdot, \cdot)$ takes diffused spectrograms $\mathbf{x}_t$ and time step $t$ as inputs, and outputs the predicted noise. By minimizing Eq.~(\ref{eq: loss}), the denoising UNet can predict the corresponding noise based on the specific $t$ and $c$. 
\begin{align}
\label{eq: loss}
    \mathcal{L} = \mathbb{E}_{t, \mathbf{x}_t, \epsilon_t}[(\epsilon_t - \mu_{\theta}(\mathbf{x}_t,t))] \ .
\end{align}

It enables gradual denoising from $\mathbf{x}_t$ through the reverse denoising process to reconstruct the spectrograms. 
\begin{align}
\label{eq: denoising_process}
    p_{\theta}(\mathbf{x}_{t-1} \, | \, \mathbf{x}_t) = \mathcal{N}(\mathbf{x}_{t-1} \, | \, \mu_{\theta}(\mathbf{x}_t,t), \ (1 - \alpha_t)\mathbf{I} ) \ ,
\end{align}
And the reconstructed $\mathbf{\tilde{x}}_{t-1}$ is calculated with the diversity coefficient $\beta_t$ and the reconstructed $\mathbf{\tilde{x}}_{0}$ is obtained iteratively: 
\begin{align}
\label{eq: stepwise_denoising}
    \mathbf{\tilde{x}}_{t-1} =  \frac{1}{\sqrt{\alpha_t}} ( \mathbf{x}_{t} - \frac{1-\alpha_t}{\sqrt{1-\bar \alpha_t}} \epsilon_{\theta}^{(t)}(\mathbf{x}_t) ) + \beta_t \epsilon_t \ ,
\end{align}

\textbf{Inference with DDIM}. While DDPM training is efficient, a major drawback of DDPM is its slow sampling speed due to the Markov chain. DDIM addresses it with a non-Markovian sampling process, which we employ to speed up inference. DDIM redefines the joint distribution $q(\mathbf{x}_{1:T} \, | \, \mathbf{x}_0)$ to enable faster and more efficient generation: 
\begin{align} 
    q_{\sigma}(\mathbf{x}_{1:T} \, | \, \mathbf{x}_0) = q_{\sigma}(\mathbf{x}_{T} \, | \, \mathbf{x}_0) \prod_{t = 2}^{T}q_{\sigma}(\mathbf{x}_{t-1} \, | \, \mathbf{x}_{t}, \mathbf{x}_0) \ ,
\end{align}
where $q_{\sigma}(\mathbf{x}_{T} \, | \, \mathbf{x}_0)$ is same as Eq. \eqref{eq: forward_process}. Based on this, the reconstructed noise-free data $\mathbf{\tilde{x}}_0$ could be predicted as follows:
\begin{align}
\label{eq: onestep_denoising}
    \mathbf{\tilde{x}}_0 
    = \frac{1}{\sqrt{\bar{\alpha}_t}} (\mathbf{x}_t - \sqrt{1 - \bar{\alpha}_t} \epsilon_\theta^{(t)}(\mathbf{x}_t, t)) \ .
\end{align}
Compared to stepwise denoising in Eq.~(\ref{eq: stepwise_denoising}), Eq.~(\ref{eq: onestep_denoising}) enables skipping-step denoising or even one-step denoising.

\subsection{Anomaly score}\label{sec:anomalyscore}
The core principle for detection is that since UD-ASD is trained only on normal data, it will attempt to reconstruct any input as a normal audio. Naturally, computing the reconstruction error ($\mathbf{x}_0 - \mathbf{\tilde{x}}_0$): normal audio yields low reconstruction error, while anomalous audio, which the model tries to reconstruct as normal, resulting in high reconstruction error. 

While simple norms like L1 or L2 can quantify this error, they often fail to capture the complex distribution of reconstruction errors and ignore crucial temporal-spectral features. The process of denoising Gaussian noise in diffusion models makes GMM well-suited for modeling the distribution of residual noise in reconstruction error. 

Getting the difference ($\mathbf{x}_0 - \mathbf{\tilde{x}}_0$), we apply temporal pooling across the time axis to capture temporal continuity, yielding a 256-dim vector. Next, a GMM with 2 mixture components (which is motivated by the bimodal distribution of machine sounds and noise) and full covariance matrices is trained on the set of error vectors derived from the normal data. During inference, the same feature extraction process is applied to a test sample. The final anomaly score is the negative log-likelihood of the sample's error vector under the trained GMM. A higher score indicates that the sample's reconstruction error pattern deviates significantly from the distribution of normal errors, and is thus classified as anomalous. This approach provides a more robust and fine-grained measure of anomaly by modeling the complete statistical distribution of normal reconstruction errors.

\section{Experiment}\label{sec:experiment}

\begin{table}[t]
    \centering
    \caption{Hyperparameters of UD-ASD.}
    \label{tab:Hyperparameters}
    \begin{tabular}{c | c | c}
        \toprule
        \multirow{2}{*}{\textbf{Diffusion}} & Diffusion steps & 1000 \\
                                            & Noise schedule & linear \\
        \cmidrule{1-3}
        \multirow{6}{*}{\textbf{UNet}} & Channels & 64 \\
                                       & Channels multiple & 1, 1, 2, 4 \\
                                       & Heads & 4 \\
                                       & Attention resolution & 32, 16, 8 \\
                                       & Dropout & 0 \\
                                       & EMA rate & 0.9999 \\
        \cmidrule{1-3}
        \multirow{4}{*}{\textbf{Training}} & Optimiser & AdamW \\
                                           & Scheduler & Cosine \\
                                           & Learning rate & 2e-4 \\
                                           & Batch size & 32 \\
        \bottomrule
    \end{tabular}
\end{table}

\begin{figure}[t]
    \centering
    \includegraphics[width=0.595\linewidth]{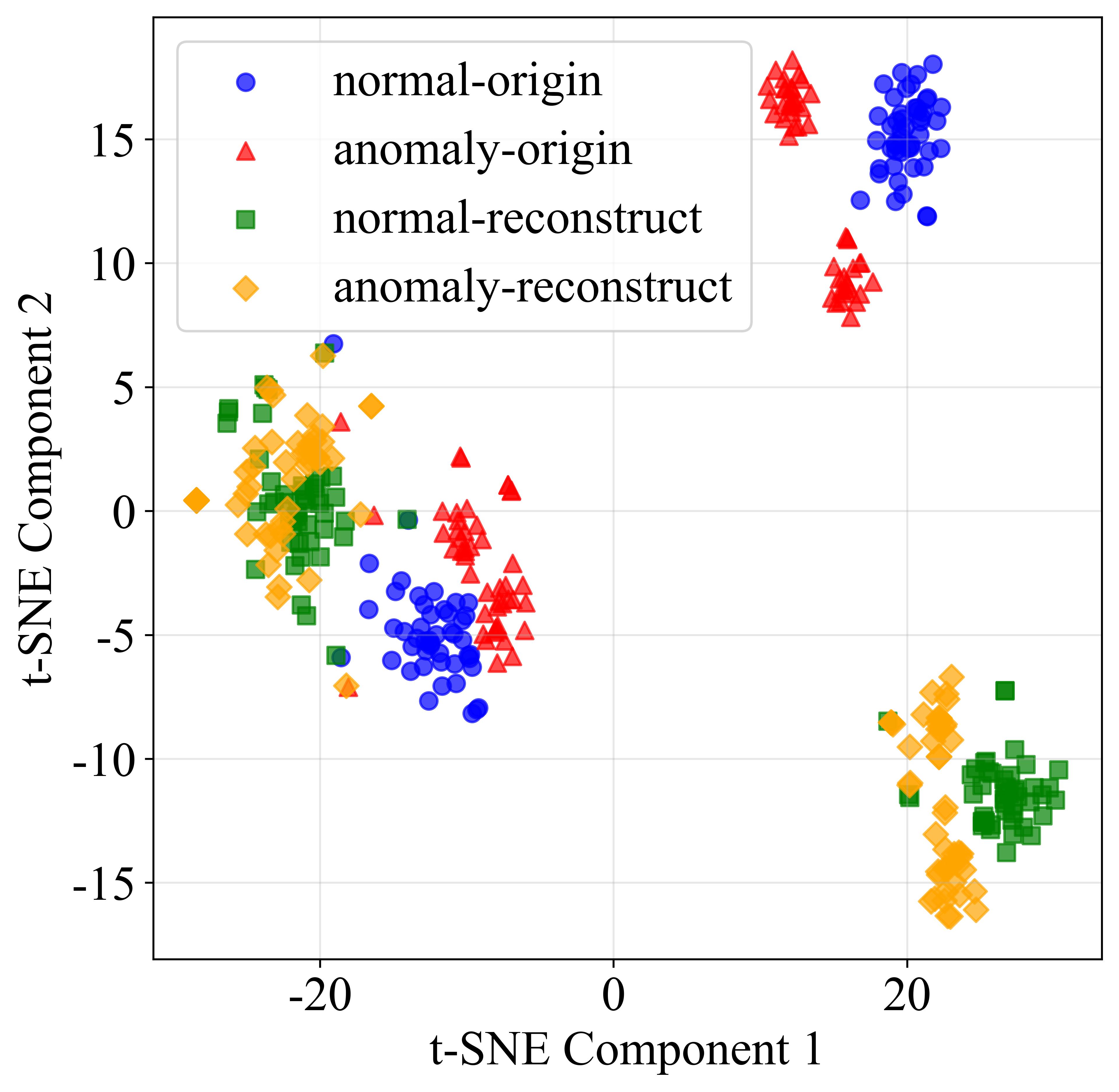}
    \caption{The t-SNE plots of Bearing Section 02. The blue and the red denote the normal original audio and the anomalous original audio, respectively. The green and the yellow represent the normal reconstruction audio and the anomalous reconstruction audio, respectively. }
    \label{fig:tsne}
\end{figure}

\subsection{Implementation}\label{sec:implementation}

\textbf{Dataset}. We evaluate UD-ASD on the DCASE2022 Challenge Task 2 development dataset \cite{dataset}. This dataset comprises audio from seven distinct machine types, each with three operational sections. The training set consists of 1,000 clips of normal operational sounds per machine section, while the test set contains a balanced mix of 50 normal samples and 50 anomalous samples. Each audio is \SI{10}{\second}.

\textbf{Implementation details}. Audio is first converted to log-Mel spectrograms with a 1024-point FFT and 256 Mel filters, then rescaled to $[-1, 1]$ in a 256$\times$256 resolution. The architecture of UD-ASD is based on \cite{beatgan}. Key hyperparameters are detailed in Table~\ref{tab:Hyperparameters}. We employ the DDPM for training and the DDIM sampler for inference acceleration with 100 steps. Random seed is 42.

\textbf{Evaluation metrics}. Following the official DCASE challenge protocol, we report the Area Under the ROC Curve (AUC) and the partial AUC (pAUC) over a low false positive rate (p=0.1). We also report the harmonic mean (Hmean) of the average AUC and pAUC scores across all machine types as the primary performance indicator.

\begin{figure*}[t]
    \centering
    \includegraphics[width=0.71\linewidth]{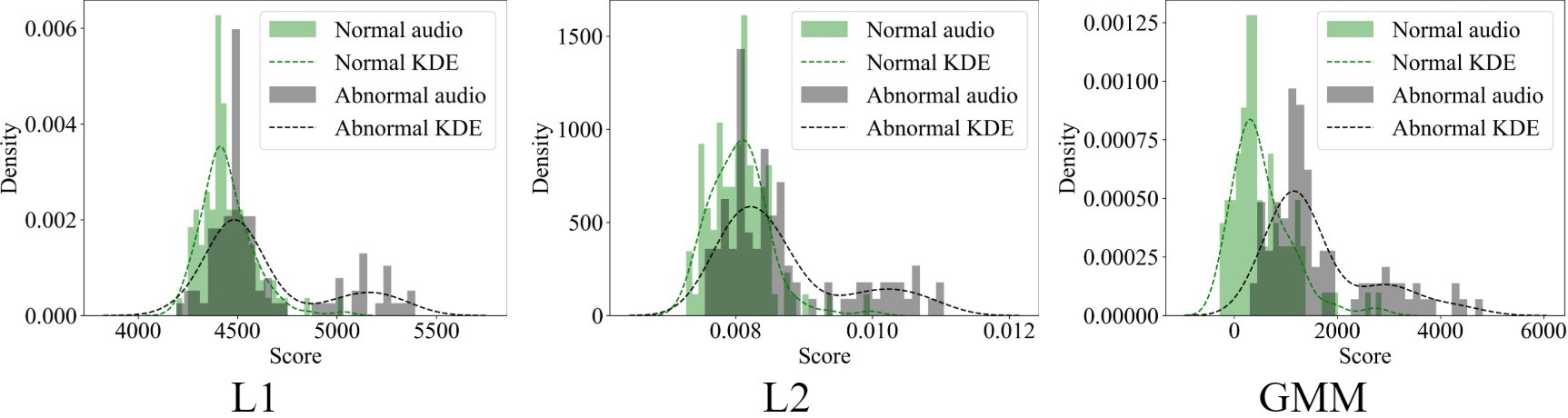}
    \caption{The histograms and KDEs for three functions on Fan Section 00, the green represents normal data and the gray means abnormal data. }
    \label{fig:ablation}
\end{figure*}

\begin{table*}[t]
\centering
\caption{Comparison with baselines and sota generative methods. Official-AE and official-CLS are the dense AE baseline and the classification baseline provided by DCASE organizers, respectively. }
\label{tab:comparison_22}
\setlength{\tabcolsep}{4pt}
\renewcommand{\arraystretch}{1.15}
\resizebox{\textwidth}{!}{
\begin{tabular}{l *{8}{cc}}
\toprule
\multirow{2}{*}{\textbf{Methods}} &
\multicolumn{2}{c}{\textbf{ToyCar}} &
\multicolumn{2}{c}{\textbf{ToyTrain}} &
\multicolumn{2}{c}{\textbf{Bearing}} &
\multicolumn{2}{c}{\textbf{Fan}} &
\multicolumn{2}{c}{\textbf{Gearbox}} &
\multicolumn{2}{c}{\textbf{Slider}} &
\multicolumn{2}{c}{\textbf{Valve}} &
\multicolumn{2}{c}{\textbf{Hmean}} \\
\cmidrule(lr){2-3}\cmidrule(lr){4-5}\cmidrule(lr){6-7}\cmidrule(lr){8-9}\cmidrule(lr){10-11}\cmidrule(lr){12-13}\cmidrule(lr){14-15}\cmidrule(l){16-17}
& \textbf{AUC} & \textbf{pAUC} & \textbf{AUC} & \textbf{pAUC} & \textbf{AUC} & \textbf{pAUC} & \textbf{AUC} & \textbf{pAUC} & \textbf{AUC} & \textbf{pAUC} & \textbf{AUC} & \textbf{pAUC} & \textbf{AUC} & \textbf{pAUC} & \textbf{AUC} & \textbf{pAUC} \\
\midrule
Official-AE             & 61.18 & 60.21 & 43.14 & 49.36 & 59.93 & 53.95 & 41.16 & 50.12 & 61.92 & 51.95 & 58.95 & 54.16 & 54.26 & 51.30 & 53.01 & 52.80 \\
Official-CLS    & 42.79 & 53.44 & 51.22 & 50.98 & 58.23 & 52.16 & 30.34 & 55.22 & 51.34 & 48.49 & 62.42 & 53.07 & 72.77 & 65.16 & 49.19 & 53.67 \\
Du\cite{Du}  & 71.57 & 54.76 & 47.67 & 50.42 & 64.04 & 52.82 & 52.72 & 57.27 & 73.32 & 60.32 & 73.98 & 62.96 & 58.47 & 50.40 & 52.32 & 55.21 \\
Yamashita\cite{Yamashita}     & 80.65 & 54.49 & 71.94 & 52.77 & 48.26 & 54.42 & 54.32 & 55.16 & 76.21 & 61.55 & 76.16 & 65.36 & 58.62 & 50.55 & 64.37 & 55.94 \\
AEGAN-AD\cite{AEGAN-AD}             & 83.26 & 69.65 & 60.00 & 52.17 & 69.34 & 60.46 & 62.47 & 59.81 & 78.54 & \textbf{69.55} & 78.30 & 66.54 & 55.46 & 52.88 & 68.20 & 60.82 \\
HMIC-AGC\cite{HMIC-lan24}             & \textbf{87.91} & \textbf{77.51} & 59.10 & 52.83 & 68.14 & 59.41 & 57.63 & 53.25 & 79.78 & 61.29 & 80.76 & 58.29 & \textbf{89.87} & \textbf{82.30} & 71.79 & 61.91 \\
DP-MAE\cite{DP-MAE}             & 79.30 & - & 66.22 & - & 76.10 & - & 64.30 & - & 73.79 & - & 83.35 & - & 79.94 & - & 74.71 & - \\
ASD-Diff\cite{ASD-Diff-fan24}             & 72.54 & - & - & - & 65.16 & - & 60.68 & - & 69.59 & - & 73.25 & - & - & - & 67.90 & - \\
MFPPG\cite{MFPPG-lu26}             & 67.25 & - & 64.18 & - & 66.96 & - & 69.06 & - & 69.31 & - & 80.58 & - & 71.23 & - & 70.14 & - \\
\midrule
UD-ASD-S             & 84.75 & 60.88 & 66.46 & 51.55 & \textbf{81.54} & \textbf{65.92} & 76.23 & 64.08 & 79.14 & 68.56 & 86.50 & 66.12 & 53.98 & 50.50 & 73.72 & 60.28 \\
UD-ASD-U             & 83.91 & 58.95 & \textbf{72.74} & \textbf{54.38} & 68.37 & 54.53 & \textbf{78.11} & \textbf{66.07} & \textbf{82.28} & 68.32 & \textbf{87.45} & \textbf{73.97} & 71.28 & 63.41 & \textbf{77.16} & \textbf{62.80} \\
\bottomrule
\end{tabular}}
\end{table*}

\subsection{Performance comparison}\label{sec:comparison}

Table~\ref{tab:comparison_22} compares our method with other models. We report two versions of our model: UD-ASD-S, which trains a separate model for each machine, and UD-ASD-U, our proposed unified model trained on all machine types simultaneously. 
Our UD-ASD-U achieves the highest overall Hmean, noting superior performance across machines. It secures the best results on four out of seven machine types. Notably, on the Fan dataset, where anomalies often manifest as continuous frequency shifts, UD-ASD-U outperforms the official-AE baseline by +36.95\% AUC. This highlights the diffusion model's superior capability in capturing and reconstructing fine-grained spectral details compared to standard AEs. While some specialized models like HMIC-AGC \cite{HMIC-lan24} excel on specific machines by incorporating extra hierarchical information, our unified model proves more robust and generalizable across the board. 

The enhanced performance of UD-ASD-U suggests that training on diverse machines acts as data augmentation, improving the model's ability to learn a more generalized representation of "normalcy". CP then effectively guides the model to reconstruct machine-specific spectrograms, combining the benefits of broad generalization with specialized generation. 

AE implicitly learns compressed representations of data through information bottlenecks, which are manifold distributions spread across lower dimensions. This compression causes AE to easily overlook subtle variations, resulting in the model learning only a coarse approximation of the manifold distribution. Diffusion, in contrast, performs explicit, fine-grained repairs. This forced “normalization” repair often becomes overly sensitive to subtle anomalous variations, making it difficult to perfectly reconstruct any data deviating from the normal distribution and thus causing greater reconstruction errors. Furthermore, the one-model-per-machine paradigm is prone to overfitting within specific domains (i.e., corresponding machine types). Cross-domain knowledge transfer, however, compels a unified model to learn more fundamental decoupled feature spaces. This acts as a powerful regularization, enhancing the model's generalization capabilities and interpretability. 

In Figure~\ref{fig:tsne}, the distinction between the blue and the red is pronounced, while the green and the yellow tend to resemble. This occurs because UD-ASD attempts to reconstruct abnormal data into normal ones, enabling the identification of anomalies through reconstruction error. In Figure~\ref{fig:tsne}, the data for each color is divided into two parts, so we use two mixture components in GMM.
Regarding efficiency, while diffusion models are known for slow inference, the DDIM sampler makes our approach practical. On an NVIDIA A40 GPU, UD-ASD (24.4M parameters) achieves an average inference time of \SI{0.32}{\second} per clip, which is well within acceptable limits for many industrial monitoring applications.
UD-ASD also faces limitations: 1) It relies on the availability of accurate machine labels, its efficacy in scenarios with missing or error labels requires further investigation. 2) UD-ASD is not designed for zero-shot detection on entirely new machine types not seen during training.

\subsection{Ablation study}\label{sec:ablation}

\textbf{Effectiveness of unified model and CP}. To validate the efficacy of our unified approach, we compare UD-ASD-U against UD-ASD-S in Table~\ref{tab:comparison_22}. While UD-ASD-S already shows competitive results, UD-ASD-U surpasses it in Hmean and on five machines. This demonstrates that CP successfully provides the necessary conditioning for the model to differentiate between machine features. More importantly, it confirms our hypothesis that a unified model can leverage cross-machine data to enhance its overall representation learning, leading to better performance than isolated training. 

\begin{table}[t]
    \centering
    \caption{Comparison with different anomaly score functions.}
    \begin{tabular}{cccc}
    \toprule
    \textbf{Functions} & \textbf{AUC source} & \textbf{AUC target} & \textbf{pAUC} \\
    \midrule
    L1             & 73.97 & 65.48 & 56.01      \\
    L2             & 74.41 & 64.18 & 55.64      \\
    GMM            & \textbf{79.83} & \textbf{73.35} & \textbf{66.82}      \\
    \bottomrule
    \end{tabular}
    \label{tab:ablation}
\end{table}

\textbf{Impact of anomaly scoring}. 
We investigate the choice of anomaly scoring by comparing GMM against standard L1 and L2 norms. As shown in Table~\ref{tab:ablation}, GMM consistently yields the best results. To visualize this, Figure~\ref{fig:ablation} plots the distributions for normal (green) and abnormal (gray) samples from the challenging Fan machine. For L1 and L2 norms, the distributions exhibit significant overlap, making it difficult to set an effective decision threshold. In contrast, GMM shows a clearer separation between the two classes. This is because the GMM models the complex, potentially multi-modal distribution of normal reconstruction errors, providing a more powerful and robust statistical measure of deviation than simple pixel-wise norms.

\section{Conclusion}\label{sec:conclusion}

We propose the unified diffusion-based model, UD-ASD, for ASD, which is only trained on normal audio, resulting in high reconstruction errors for abnormal audio. The multistep diffusion and denoising process gives UD-ASD strong generative capabilities. Condition Projector enables a single unified model to handle multiple machine types, significantly reducing computational costs. Experiments on the DCASE2022 Task 2 dataset confirm the effectiveness of UD-ASD.

\vfill\pagebreak

\section{Generative AI Use Disclosure}
We use generative AI solely to help polish our paper. And we do not use generative AI to complete any part of the work on its own.

\bibliographystyle{IEEEtran}
\bibliography{mybib}

\end{document}